# Uncertainty in Self-Adaptive Systems: A Research Community Perspective

SARA M HEZAVEHI, University of Groningen, The Netherlands, Linnaeus University, Sweden
DANNY WEYNS, Katholieke Universiteit Leuven, Belgium, Linnaeus University, Sweden
PARIS AVGERIOU, University of Groningen, The Netherlands
RADU CALINESCU, University of York, UK
RAFFAELA MIRANDOLA, Politecnico di Milano, Italy
DIEGO PEREZ-PALACIN, Linnaeus University, Sweden

One of the primary drivers for self-adaptation is ensuring that systems achieve their goals regardless of the uncertainties they face during operation. Nevertheless, the concept of uncertainty in self-adaptive systems is still insufficiently understood. Several taxonomies of uncertainty have been proposed, and a substantial body of work exists on methods to tame uncertainty. Yet, these taxonomies and methods do not fully convey the research community's perception on what constitutes uncertainty in self-adaptive systems, and on the key characteristics of the approaches needed to tackle uncertainty. To understand this perception and learn from it, we conducted a survey comprising two complementary stages in which we collected the views of 54 and 51 participants, respectively. In the first stage, we focused on current research and development, exploring how the concept of uncertainty is understood in the community, and how uncertainty is currently handled in the engineering of self-adaptive systems. In the second stage, we focused on directions for future research, to identify potential approaches to dealing with unanticipated changes and other open challenges in handling uncertainty in self-adaptive systems. The key findings of the first stage are: a) an overview of uncertainty sources considered in self-adaptive systems, b) an overview of existing methods used to tackle uncertainty in concrete applications, c) insights into the impact of uncertainty on non-functional requirements, d) insights into different opinions in the perception of uncertainty within the community, and the need for standardised uncertainty-handling processes to facilitate uncertainty management in self-adaptive systems. The key findings of the second stage are: a) the insight that over 70% of the participants believe that self-adaptive systems can be engineered to cope with unanticipated change, b) a set of potential approaches for dealing with unanticipated change, c) a set of open challenges in mitigating uncertainty in self-adaptive systems, in particular in those with safety-critical requirements. From these findings, we outline an initial reference process to manage uncertainty in self-adaptive systems. We anticipate that the insights on uncertainty obtained from the community, and our proposed reference process will inspire valuable future research on self-adaptive systems.

Authors' addresses: Sara M Hezavehi, University of Groningen, The Netherlands, Linnaeus University, Sweden, hezavehisara@gmail.com; Danny Weyns, Katholieke Universiteit Leuven, Belgium, Linnaeus University, Sweden, danny.weyns@kuleuven.be; Paris Avgeriou, University of Groningen, The Netherlands, p.avgeriou@rug.nl; Radu Calinescu, University of York, UK, radu.calinescu@york.ac.uk; Raffaela Mirandola, Politecnico di Milano, Italy, raffaela.mirandola@polimi.it; Diego Perez-Palacin, Linnaeus University, Sweden, diego.perez@lnu.se.

## 1 INTRODUCTION

Self-adaptation was introduced about two decades ago as a means to manage the growing complexity of computing systems [15, 21]. Multiple terms have been used to refer to systems with self-adaptation capabilities, including *autonomic systems* [15], *organic computing systems* [27], *dynamic adaptive systems* [37], and *self-adaptive systems* [12]. We use the last term in this paper. While the initial focus of the research in this area was on automating the complex task of system operators, about a decade ago researchers and engineers started to realise that the presence of uncertainty is a central aspect of self-adaptation [11, 34].

Self-adaptation is achieved by enhancing a system with a feedback loop (or a combination of feedback loops) [6, 26, 33]. The task of this feedback loop is to ensure that the system complies with a set of adaptation goals when the operating conditions change. Adaptation goals are typically non-functional properties of the system, such as performance, reliability, and cost [35]. To that end, the feedback loop *monitors* the managed system (i.e., the system that is subject to adaptation) and its environment to collect runtime data that were not available before, *analyses* and *plans* alternative configurations, and *executes* adaptations to achieve the adaptation goals, or to degrade gracefully if needed. The monitor-analyse-plan-execute functions and the *knowledge* underpinning their operation are often referred to as a MAPE-K loop [15].

Self-adaptation blurs the traditional divide between the offline system development activities performed by engineers (supported by tools) and the online activities performed by the system (possibly supported by human stakeholders) [1, 2]. As such, a self-adaptive system can be viewed as a temporally completed system with some degrees of freedom in terms of its configuration. This allows the self-adaptive system to select or adapt its configuration when the conditions change, and thus to deal with uncertainties that are difficult or impossible to anticipate before deployment. At runtime, the system collects additional information to resolve the uncertainty and to adapt itself in order to preserve its adaptation goals under changing conditions.

Unfortunately, uncertainty is a complex concept that is difficult to understand, let alone to manage. Given the central role of uncertainty in self-adaptive systems, substantial efforts have been devoted to "taming" uncertainty in the past years, with a particular emphasis on devising taxonomies of uncertainty for self-adaptive systems [9, 18, 22, 24] and on developing methods for managing uncertainty, e.g., [4, 7, 8, 19]. While these taxonomies and methods have been instrumental in putting the focus on uncertainty as a key driver for self-adaptation, they do not fully convey the perception of the research community on what constitutes uncertainty in self-adaptive systems, and what key characteristics must be considered by the approaches that aim at tackling uncertainty.

To understand this perception and learn from it, we conducted a survey comprising two complementary stages. In the first stage, we focused on current research and development, exploring how the concept of uncertainty is understood by the research community, and how uncertainty is handled in the engineering of concrete self-adaptive systems. In the second stage, we focused on directions for future research, identifying insights on the ability of self-adaptive systems to deal with unanticipated changes and potential approaches for dealing with them, and open challenges in handling uncertainty in self-adaptive systems. All survey participants (54 and 51 for stages 1 and 2, respectively) are actively involved in research on self-adaptation in the broader community.[1]

This paper presents an extensive analysis of the data from our survey, and provides the following contributions: (i) an overview of uncertainty sources considered in self-adaptive systems, (ii) an overview of existing methods used to tackle uncertainty in concrete applications, (iii) insights into the impact of uncertainty on non-functional requirements, (iv) insights into different opinions in the perception of uncertainty within the research community, (v) insights into the beliefs of the community that self-adaptive systems can cope with unanticipated change, (vi) a set of potential approaches for dealing with unanticipated change, and (vii) a set of open challenges in mitigating uncertainty in self-adaptive systems, in particular in those with safety-critical requirements. Based on the results of our study, we outline an initial reference process for uncertainty management in self-adaptive systems. This process defines a set of coordinated activities that span the different stages of the lifecycle of a self-adaptive system.

---

[1]We use the term 'broader community' to refer to the self-adaptive systems researchers who meet regularly at the International Symposium on Software Engineering for Adaptive and Self-Managing Systems (SEAMS), the International Conference on Autonomic Computing and Self-Organising Systems (ACSOS, ICAC and SASO), and other related events.

The paper extends significantly our earlier work [3] that presented preliminary results on the research community's views on the ability of self-adaptive systems to deal with unanticipated change, and on open challenges in mitigating uncertainties in self-adaptive systems. The main extensions of our previous work include novel results on: (i) uncertainty sources in self-adaptive systems, (ii) existing methods used to tackle uncertainties, (iii) the impact of uncertainty on non-functional requirements, and (iv) inconsistencies in the perception of uncertainty within the research community. In addition, we exploit the results obtained from the integrated survey to outline an initial uncertainty management process for self-adaptive systems, responding to the need for standardised uncertainty-handling processes that our study has identified.

The remainder of this paper is organised as follows. In Section 2, we summarise related work, in particular taxonomies for uncertainty in self-adaptive systems and selected approaches for handling uncertainty. Section 3 provides an overview of the scientific method we used in this research. We then report the analysis of the data collected in the two stages of the survey in Section 4. Next, we discuss the main findings obtained from the survey, and we outline the initial reference model for uncertainty management in self-adaptive systems in Section 5. Finally, we discuss threats to validity in Section 6, and wrap up with an outlook on future research in self-adaptive systems in Section 7.

## 2 RELATED WORK

The notion of uncertainty has been studied in a wide variety of fields, often in connection with decision-making; a recent example is [14]. A common assumption in such studies is that decision-making processes partly rely on humans. However, in self-adaptive systems the dependency of decision-making processes on humans are typically minimised as these systems are expected to operate largely autonomously, implying that their decisions are primarily made by software. This requires a fresh and innovative perspective on the problem of decision-making under uncertainty. In this section, we discuss related work on uncertainty in self-adaptive systems. We start with a set of studies that provide basic insights in the notion of uncertainty in self-adaptive systems. Then, we discuss a number of taxonomies and classifications of uncertainty in adaptive systems. Next, we highlight a selection of representative studies that apply concrete techniques to tame uncertainty. Finally, we summarise the current state of research and motivate the study presented in this paper.

*Basics of Uncertainty in Self-Adaptive Systems*: Back in 2010, Garlan [11] already highlighted the key role of uncertainty management in software-intensive systems. He discussed several sources of uncertainty affecting modern software systems: humans in the loop, learning, mobility, cyber-physical systems, rapid evolution, and argued that uncertainty in software systems should be considered as a first-class concern throughout the whole system lifecycle. Esfahani and Malek [9] studied uncertainty in self-adaptive systems with an emphasis on sources of uncertainty that include: simplifying assumptions, model drift, noise, parameters in future operation, human in the loop, objectives, decentralization, context, and cyber-physical systems. That study also investigated uncertainty characteristics (reducibility versus irreducibility, variability versus lack of knowledge, and spectrum of uncertainty). In [34], Weyns provided a perspective on the evolution of the field of self-adaptation in seven waves of research that span the last two decades. The fifth wave, "Guarantees Under Uncertainties", defines uncertainty as a central driver for self-adaptation. That work classifies uncertainty sources into four major groups: uncertainty related to the system itself, uncertainty related to the system goals, uncertainty in the execution context, and uncertainty related to human aspects, and explores how these sources can be handled using various approaches.

The work on the basics of uncertainty in self-adaptive systems summarised above has been pivotal in creating understanding in the key role of uncertainty in self-adaptation. These insights primarily originate from literature material complemented with personal insights from individual experts. The work presented in this paper complements these insights with the research community's perception on what constitutes uncertainty in self-adaptive systems, and how to tackle uncertainty.

*Taxonomies and Classifications*: Ramirez et al. [24] provided a taxonomy for uncertainty in dynamically adaptive systems. The taxonomy classifies sources of uncertainty for the requirements, design, and runtime phases of dynamically adaptive systems. The uncertainties are described using a template inspired by the established template for representing design patterns (name, classification, context, impact, mitigation strategies, sample illustration, related sources). Perez-Palacin et al. [22] presented a taxonomy for uncertainty in self-adaptive systems modeling that comprises three key dimensions: location, level, and nature. The location of uncertainty refers to the model aspects affected by the uncertainty. The level of uncertainty indicates where the uncertainty is placed on the spectrum between deterministic knowledge and total ignorance. Finally, the nature of uncertainty shows whether the uncertainty is due to the imperfection of the acquired knowledge or to the inherent variability of the modelled phenomena. Mahdavi et al. [18] reviewed uncertainty in self-adaptive systems with multiple requirements. From the data collected from 51 primary studies, the authors derive a systematic overview of uncertainty dimensions (location, nature, level/spectrum, emerging time, sources) with their respective options. The sources of uncertainty are further elaborated and are grouped into several classes, i.e., uncertainty of: models, adaptation functions, goals, environment, resources, and managed system. Musil et al. [20] introduced a number of uncertainty types and argued for their significant role in Cyber Physical Systems (CPS) based on data extracted from a systematic mapping study. The authors explored three main paradigms for realising adaptation in CPS, namely, architecture-based adaptation, multi-agent based approaches, and self-organising based approaches. More recently, Troya et al. [31] presented the results of a survey focused on the uncertainty representation in software models. Specifically, the authors identified existing notations and formalisms used to represent the different types of uncertainty, together with the software development phase in which they are used and the types of analysis allowed. To analyse the state of the art, a classification framework is introduced that allows the comparison, classification, and analysis of existing approaches. The study highlights the need to move the maturity level in this area from the initial stage focused on scientific studies to a more advanced one in which it is possible to influence and improve the industrial modelling practices.

The work on taxonomies and classifications has contributed to deeper insights in the sources, nature, and representation of uncertainty in self-adaptive systems. Most work is grounded in empirical evidence. Yet, the results are derived from material documented in research papers. Our study complements these results with the insights derived from the perception of active members of the community elicited though a two-stage survey.

*Selection of Concrete Approaches*: Over the past years, the number of studies on adaptation approaches that take into account uncertainty as first-class citizen in self-adaptive systems has gradually been increasing. We discuss a representative sample of this work. Moreno et al. [19] proposed a method for improving decision–making in self-adaptive system driven by reducing uncertainty. The authors list uncertainty types associated with different sources and discuss when uncertainty should be reduced. Then they present examples of uncertainty reduction tactics and different activities in a typical self-adaptation loop. In their work, Camara et al. [5] explored existing techniques for uncertainty management in self-adaptive systems. The authors then presented a method to represent different types of uncertainty within the MAPE-K framework and offer techniques to mitigate specific types of uncertainty. In [16] Kinneer et al. addressed the problem of handling changes to the self-adaptive system itself, such as the addition or removal of adaptation tactics. To avoid human planning, they proposed reuse of prior planning knowledge using genetic programming. The focus is on investigating a planner's ability to address unexpected adaptation needs through plan reuse. In [28] Shevtsov et al. presented SimCA*, a control-theoretic approach to handle uncertainty in self-adaptive software systems. The authors consider different types of uncertainty: disturbances originating from the environment such as noise, changes of system parameters, and predefined changes of quality requirements at runtime.

As illustrated with this selection of work, progress in the field is taking place. Yet, beyond concrete contributions, research efforts are required that aim at a more holistic understanding of uncertainty in self-adaptive systems. This is the overall aim of the study presented in this paper.

*Summary:* A body or work exists on uncertainty in self-adaptive systems. This work has studied the notion of uncertainty and mechanisms to mitigate uncertainty based on existing research literature and individual projects and experiences. Our paper complements these important efforts by presenting insights into the research community's *perception* of the notion of uncertainty. To that end, we asked experts from the broader community of self-adaptive systems to express their view on: what uncertainty is and how it can be handled in practical applications, whether unanticipated changes can be handled and (if so) by what means, and which open challenges still limit the ability of self-adaptive systems to handle uncertainty. As detailed in the rest of this paper, the analysis of their responses provides unique insights into the current state of the art in dealing with uncertainty in self-adaptive systems, and highlights important potential areas for further improvements.

## 3 SCIENTIFIC APPROACH

We defined the goal of this research using the Goal-Question-Metric (GQM) approach following [32]:

**Analyze** self-adaptive systems **for the purpose of** investigation and characterization **with respect to** the concept and sources of uncertainty, as well as approaches used to handle uncertainty **from the point of view of** researchers with expertise in the engineering of prototypes.

To tackle this high-level goal we followed the scientific approach illustrated in Figure 1. The common goal is refined in two sets of complementary research questions. These sets of questions are tackled in two stages each comprising a corresponding survey. The collected data is then analysed and presented to answer the research questions and derive conclusions.

The first stage of the study focused on current research and development of uncertainty management in self-adaptive systems (SAS) as perceived by experts of the research community. The aim of the first stage is to obtain insights in the notion of uncertainty, the sources of uncertainty, methods for handling uncertainty and their impact on non-functional requirements (NFRs). The second stage focused on directions for future research as perceived by experts of the self-adaptive system's community. The aim of the second stage is to shed light on the ability of self-adaptive systems in dealing with unanticipated changes and potential methods to deal with such changes, and the challenges of tackling uncertainty in self-adaptive systems, with a particular focus on systems with strict goals. These insights can help to identify commonalities in how uncertainty is understood by the community but also potential inconsistencies in the viewpoints, better understand how uncertainty is treated in the engineering of self-adaptive systems, and identify limitations of existing methods applied to applications. As such, the results of this research help to understand whether existing methods are considered as sufficient or not, pinpoint areas for improvement, and outline key open challenges as perceived by experts in the area.

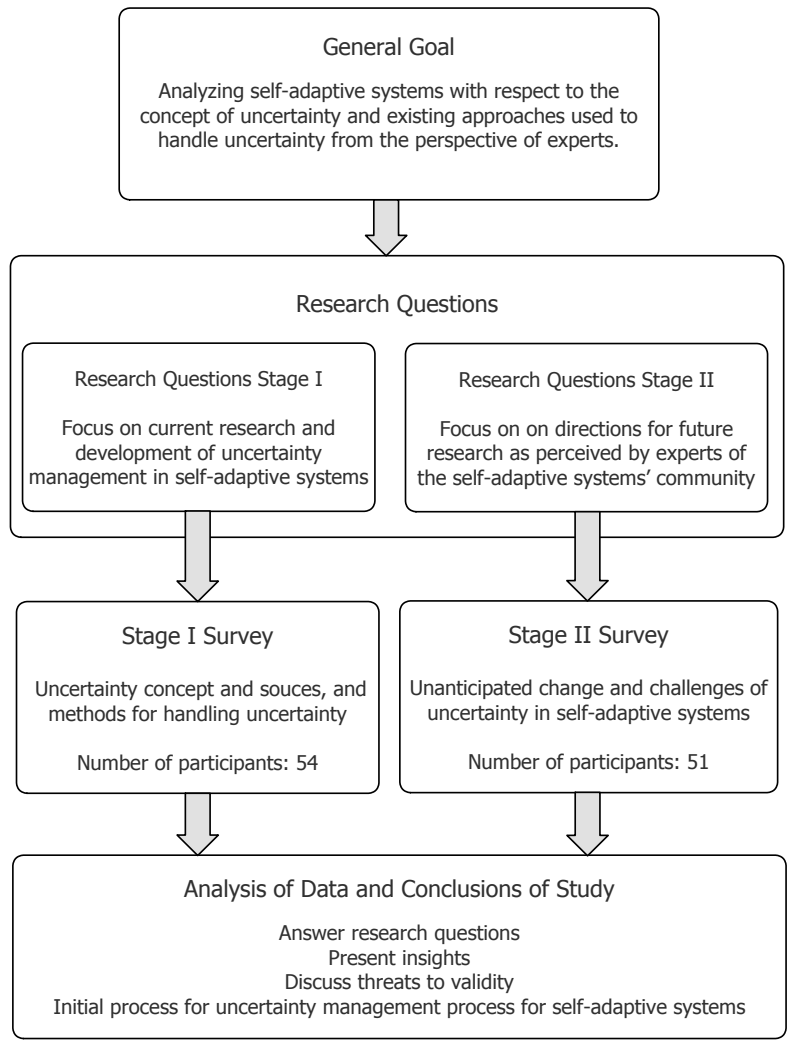

Fig. 1. Overview of the scientific approach used in this research.

The choice for the two-staged research approach is motivated qualitatively and pragmatically. By organising the data collection in two stages we could focus the questionnaires on two well-defined and coherent topic areas: current research and development on uncertainty on the one hand, and challenges for future research on uncertainty on the other hand. Focusing each survey on one topic area reduces and eases the data collection. Furthermore, surveys with many questions are tedious for respondents to fill out, and the data quality may consequently decline [36].

In the following subsections, we start with the formulation of the research questions for both stages of the research. Then we describe how we designed and conducted the survey for both stages to answer the research questions. All the data of the research presented in this paper, including the questionnaires and all responses, is available online in a replication package [13].

## 3.1 Research Questions

The high-level goal described above was refined in two sets of research questions each mapping to one stage of the research, see the lists in Table 1.

Table 1. List of research questions for the two-staged study.

| Research Question | Purpose |
| --- | --- |
| **Stage One: Current Research and Development of Uncertainty in SAS** | |
| **RQ1:** What is the perception of experts on uncertainty in SAS? | To better understand how experts define uncertainty and its sources in SAS. |
| **RQ2:** What approaches do experts use to mitigate uncertainty and deal with specific sources of uncertainty in applications? | To identify and investigate: 1) sources of uncertainty, 2) common approaches used to resolve uncertainty, 3) whether approaches are used at design- or run-time, 4) whether uncertainty is handled in a systematic or ad-hoc manner. |
| **RQ3:** How do existing approaches tackle uncertainties that affect NFRs and their trade-offs? | To investigate: 1) whether or not NFRs and their trade-offs are taken into account, and 2) if/how evidence for system compliance with the NFRs after adaptation is supported. |
| **Stage Two: Unanticipated Change and Open Challenges of Uncertainty in SAS** | |
| **RQ4:** What is the perception of the community on the ability of self-adaptive systems to deal with changes that were not anticipated when the systems were engineered? | To gain insight into the scope of uncertainty, i.e., the extent to which a self-adaptive system can handle changes that were not anticipated before system deployment. |
| **RQ5:** What are the challenges for uncertainty in self-adaptive systems perceived by the community, in particular for systems with strict goals? | To gain insight into the main challenges researchers see with respect to dealing with uncertainty, in particular for systems with goals that are critical to stakeholders. |

## 3.2 STAGE ONE - Current Research and Development of Uncertainty in SAS

Figure 2 gives an overview of the main activities we performed to design and conduct the stage one survey. Following the established guidelines of [17], we defined four main activities: creating and testing the questionnaire, finding the sample, contacting the participants, and analysing the collected data and reporting the survey results. The protocol of the stage one survey was devised and refined iteratively by three of the researchers involved in this research.

*3.2.1 Creating the Questionnaire.* To create the questionnaire, we devised a set of questions starting from research questions RQ1-RQ3. The questionnaire questions aimed at collecting accurate data from the participants to facilitate data analysis and answer the research questions. The questions focused on the concept of uncertainty (3 questions), methods for mitigating uncertainty (4 questions), and handling NFRs in the presence of uncertainty (2 questions), respectively. Eight of questions of the semi-structured questionnaire were a combination of both open and closed-ended, and one question was closed-ended. The open-ended questions were meant to provide the participants with the opportunity to freely articulate their perceptions and experiences regarding uncertainty in SAS. For a list of questions of the questionnaire with motivations, we refer to replication package [13].

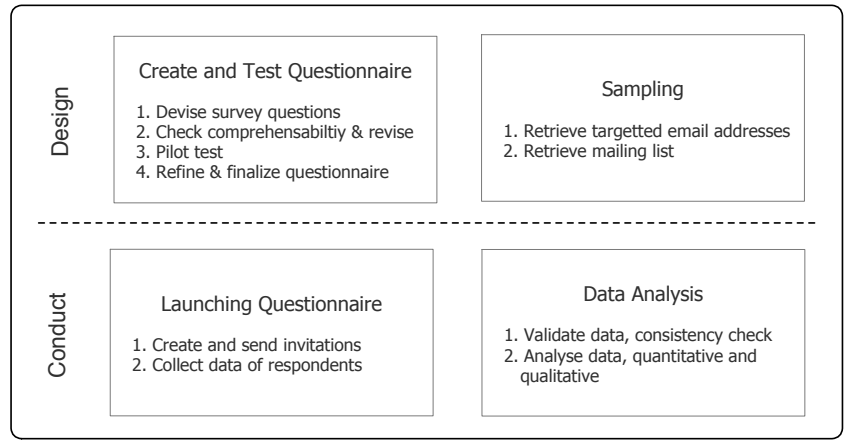

Fig. 2. Phases and activities of designing and conducting the stage one survey.

*3.2.2 Testing the Questionnaire.* We checked the comprehensibility of the questions with a number of researchers. In particular, we evaluated the questionnaire on three aspects: a) validity, b) completeness of the questionnaire, and c) clarity.

Validity refers to technical validity and usefulness of the questions. In order to examine the technical validity of the questions and inspect whether or not the questionnaire would result in the collection of the required data, i.e., usefulness, to fulfill the goal of the survey (i.e., answering the main research questions), we performed a first pilot to evaluate the questions. To conduct the test, we asked five academics (i.e. three senior researchers and two junior researchers) experienced with self-adaptive and autonomous systems to review our questionnaire and provide feedback.

The second and third aspects (i.e., completeness and clarity) helped us to investigate whether or not the questions are comprehensible and whether the questionnaire is complete, or we are missing important questions. To that end, we performed a second pilot by inviting three researchers (i.e., one senior member, and two PhD students) with backgrounds in collective adaptive systems, systems of systems, and software architecture and distributed systems to take part in the pilot to identify potential points of ambiguity in the questions. As the emphasis of this pilot was on the formulation and understandability of the questions, the experience of the participants with engineering SAS was not a primary concern.

The participants of the two pilots expressed interest in the final results and findings. We received several recommendations to improve the questionnaire. Based on this feedback, we merged a few questions, rephrased and clarified certain ambiguous points, and finalised the questionnaire.

*3.2.3 Sampling.* To create our sample, i.e., the target audience for data collection, we used non-probabilistic methods. Non-probabilistic samples are created when participants are believed to be representative of the population and the target population is very specific and of limited availability [17]. We specifically selected the convenience sampling method (i.e., a type of non-probabilistic sampling method which aims at reaching out to a group of people who are relatively easy to contact) in order to obtain respondents who are available and willing to participate our survey. Specifically, we reached out to experts[2] who are experienced in the engineering of SAS.

Concretely, to send invitations to the potential participants, we used a combination of a targeted and generalised approach. For the targeted approach, the selection of invitees was based on the significance and relevance of their published research: we selected subjects whose work is considered prominent with respect to SAS and they are most likely familiar with uncertainty relevant concepts and methods to handle it. We retrieved the email addresses of these researchers, and invited them through a personalised email. For the generalised approach, we created and sent out emails to the Software Engineering for Adaptive & Self-Managing Systems mailing list (icse-seams@sigsoft.org) inviting a wider range of researchers to participate in the survey (i.e., generalised approach).

---

[2]With "expert" we refer to researchers who are experienced in the design, development, and implementation of SAS.

*3.2.4 Launching the Questionnaire.* We prepared an online survey and contacted the sample by sending them invitations to participate in the survey. We used an online questionnaire to collect data from the participants, consisting of three main sections that correspond to the three research questions. On average, it took 20 minutes for participants to complete the questionnaire. One of the researchers involved in this study collected the data into a spreadsheet for further analysis.

*3.2.5 Data Analysis.* To perform data analysis, first we tabulated the extracted raw data in excel files for analysis purposes. The data set consisted of both quantitative data (i.e., responses based on pre-defined categories) and qualitative data (i.e., free-text answers to open-ended questions).

(1) *Quantitative data analysis.* To analyze the quantitative data, we used descriptive statistics to present quantitative descriptions for the extracted data, and to summarize the data in a comprehensible and manageable format to answer the research questions. We used descriptive statistics (e.g., mean, standard deviation, correlation, etc.), where applicable, to represent results in simpler format and combined them with plot representations of the analyzed data to answer the research questions.

(2) *Qualitative data analysis.* For the analysis of the qualitative data, we used constant comparative analysis [30]. This method offers the means to analyze the textual responses to open-ended questions. This analysis entails deriving categories and the relationship between the categories through inductive reasoning. This allowed us to investigate the possibility of integrating the participants' responses into a model explaining how uncertainty is being handled in SAS in research practice. In addition, we used coding [30] to capture the essence of data and arranging them in a systematic order for further synthesis, and then cross-checked them with two other researchers where necessary.

## 3.3 STAGE TWO - Unanticipated Change and Open Challenges of Uncertainty in SAS

Figure 3 gives an overview of the main activities we performed to design and conduct the stage two survey. For this stage, we used a cross-sectional survey [17] with a questionnaire that we delivered to the participants personally or by email. The design of the survey followed a similar process as the stage one survey.

*3.3.1 Creating the Questionnaire.* We devised a set of questions for the questionnaire starting from research questions RQ4 and RQ5. To answer RQ4, we formulated two questions; one aimed at understanding the perception of the respondents on the ability of self-adaptive systems to deal with unanticipated changes; and the other one to gain insight into how the system may be able to gain awareness of change that it was not engineered for. To answer RQ5, we formulated four questions. With the first of these questions, we aimed at understanding the perception of the respondents on the aspects of SAS runtime models that can be associated with uncertainties. The following three questions then zoomed in on uncertainties in model parameters, the model structure, and the modeling formalism. Finally, to answer RQ5, we formulated a last question that aimed at gaining insight into the perception of the respondents on open challenges in handling uncertainty in self-adaptive systems with strict requirements.

The questionnaire combined: (i) closed questions with one or more choices complemented with a text box where respondents could elaborate on their choice using free text; and (ii) open questions that respondents could answer with free text. All questions were optional. For a complete list of questions of the questionnaire, we refer to the replication package [13].

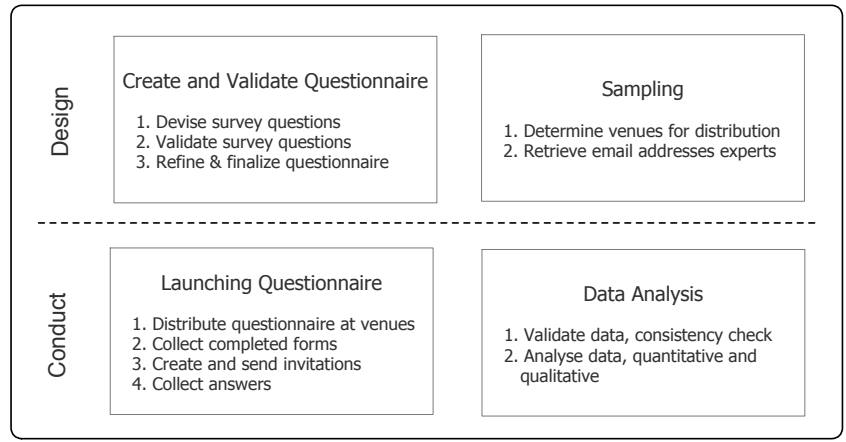

Fig. 3. Phases and activities of designing and conducting the stage two survey.

*3.3.2 Validity of the Questionnaire.* The questionnaire has been designed over several iterations based on an internal validation process. An initial set of questions was defined by three researchers in a face-to-face meeting. A fourth author then checked the questionnaire, and proposed a number of refinements plus an additional question. The revised questions were then discussed among the four researchers. After several adjustments, the questionnaire was finalised for release.

*3.3.3 Sampling.* We create our sample and collect data, we used a combination of direct and indirect methods [25]. In particular, we distributed the questionnaire to the attendees of the 2019 editions of the main venues of the SAS research community: the International Symposium on Software Engineering for Adaptive and Self-Managing Systems (SEAMS), the International Conference on Autonomic Computing (ICAC) and the International Conference on Self-Adaptive and Self-Organizing Systems (SASO).[3] Additionally, we distributed the questionnaire to the participants at the Shonan seminar on "Controlled Adaptation of Self-adaptive Systems" (CASaS) in January 2020. Each of these events was attended by at least one of the authors. To enhance validity, we have sent personally invitations via email to several additional experts of the community, inviting them to complete the questionnaire. All respondents were researchers with experience in dealing with uncertainty in self-adaptive systems. The sample included PhD students, postdoctoral researchers, and academics ranging from assistant professor to full professor.

*3.3.4 Launching the Survey.* We prepared printed copies of the two-page questionnaire to collect the data. These prints were distributed at the venues mentioned above. The respondents completed the surveys by hand. The respondents could hand in completed surveys via a box where we collected them at the end of the events. For the additional participants that we invited via email, we provided a text file that the participants could use to provide answers. The completed questionnaires were anonymously collected in a folder. One of the survey authors then copied all the answers into a spreadsheet for analysis.

*3.3.5 Data Analysis.* To analyze the data collected from the answers, we used simple descriptive statistics. In particular, for each question, we determined the percentages of the different response options. We then complemented these results by analysing the comments provided by the respondents. To that end, we applied simple qualitative data analysis using coding. This type of analysis enables identifying patterns and relationships between the data [25, 29]. The coding was performed using the following steps:

(1) *Extracting data*: we read and examined the data from the questions that allowed comments, and the answers to the open questions.
(2) *Coding data*: we did not define any coding upfront; instead we analyzed the data and incrementally added codes to small coherent fragments of the text provided in different answers (as suggested in [23]).
(3) *Translating codes into categories*: starting from the codes we then derived categories through an abstraction step where the different codes were thematically grouped.

---

[3]After the 2019 editions, ICAC and SASO merged into ACSOS.

To avoid bias in the identification of codes and the synthesis in categories, we performed both steps for each question in a team of two researchers. Both researchers worked independently and then exchanged their results. Differences where then discussed until consensus was reached. Finally, the other researchers crosschecked the results to finalize the coding.

## 4 DATA ANALYSIS AND RESULTS

We report now the analysis of the data collected from the two surveys to answer the respective research questions. The results for each research question are presented for the data that was collected for the corresponding survey.[4]

### 4.1 STAGE ONE - Current Research and Development of Uncertainty in SAS

For the survey of stage one we collected data of 53 participants. This data enables us to formulate answers to the first three research questions: RQ1 looks into the perception of the community on the concept of uncertainty (Subsection 4.1.1), RQ2 focuses on the mitigation of uncertainty (Subsection 4.1.2), and RQ3 looks at the impact of uncertainty handling methods on non-functional requirements (Subsection 4.1.3).

*4.1.1 Concept of Uncertainty.* We formulated three survey questions[5] to collect data for answering the first research question that focuses on the concept of uncertainty, see Table 2.

Table 2. Survey questions for RQ1.

| ID    | SQs for RQ1 focus on the concept of uncertainty |
|-------|---|
| SQ1.1 | Agree or disagree: "If uncertainty did not exist, there would be no need for SAS." |
| SQ1.2 | How do you define uncertainty in the context of SAS? What are the two most common SoU encountered when developing/evaluating SAS? |
| SQ1.3 | Agree or disagree: "SAS tend to face certain SoU more often than others." If agree, which types of sources? Please explain your answer. |

**SQ1.1: Uncertainty as the reason for existence of self-adaptation.**

The data of the answers to SQ1.1 is summarised in Figure 4. Sixteen participants (37%) agree that without uncertainty there is no need to apply self-adaptation. However, a majority of 34 participants (63%) do not agree that uncertainty is the only reason for applying self-adaptation. Among these, 16 do not provide any additional explanation, while six participants state that there are other motivations to apply self-adaptation independently of uncertainty but they do not clarify what these motivations actually are. The remaining 12 participants provide an explicit motivation for why uncertainty is not the only motivation to apply self-adaptation. Among these, the most frequently stated motivation for applying self-adaptation (by four participants) is dealing with variability. Variability refers to elements or parameters in the system, environment, or quality of service. Other motivations are handling internal errors of the system, dealing with complexity of the system and its domain, dealing with anticipated changes, and means to create autonomous systems (each mentioned by two participants). Two participants stated that their answer would depend on how uncertainty is defined, in particular whether or not it refers to unanticipated changes. We investigate this further in stage two of our research. Note that the answers of these two participants are not reflected in the statistical analysis of SQ1.1.

---
[4]We use SAS as abbreviation for self-adaptive systems and SoU for sources of uncertainty.
[5]We use the following notation: survey question SQx.i is the i-th survey question of research question RQx.

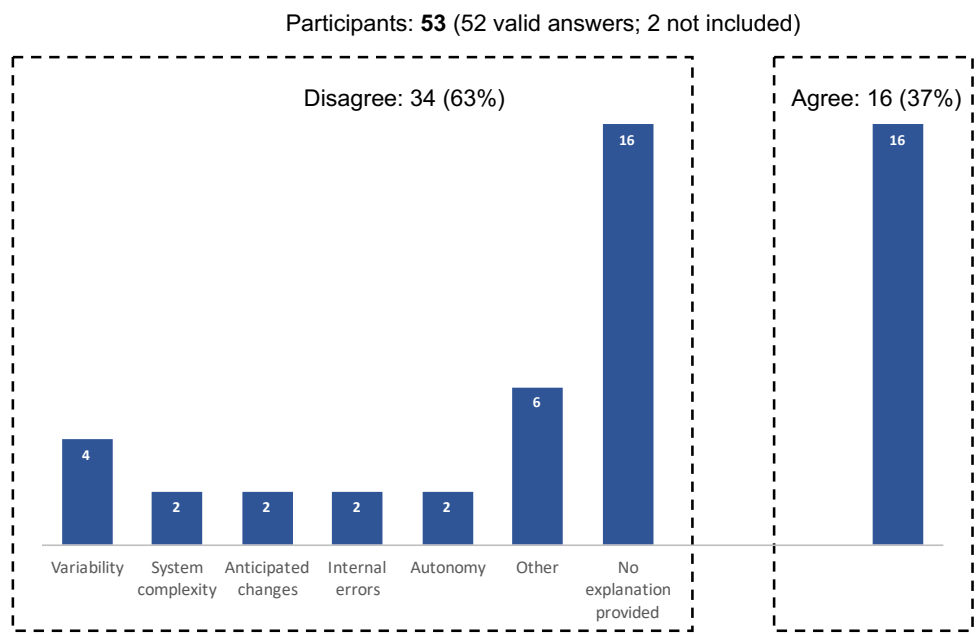

Fig. 4. Motivations other than uncertainty for applying self-adaptation.

> **Key findings from SQ1.1:** The results show that one on three participants agree that the existence of uncertainty motivates the need for self-adaptation. However, a majority of the participants state that even if uncertainty was not an issue, systems with self-adaptation capabilities would still be required. Yet, besides variability, no other major motivation for the need of self-adaptation was expressed.

*SQ1.2: Definition of Uncertainty and Common Sources of Uncertainty.*

Table 3 summarises the definitions of uncertainty that appeared most frequently. Twenty-five participants (47%) define uncertainty as (partial or complete) lack of knowledge regarding elements of the system and the environment at design time. The second most common way to define uncertainty (provided by 11 participants, i.e., 21%) is unpredictable situations (both internal to the system and and external) which the system needs to deal with at run-time. Internal situations refer to unpredictability of system components such as failures, while external situations refer to unforeseeable environmental situations concerning unavailability of resources or changes in the system's operational environment. Other relevant definitions of uncertainty given by participants are a property of the environment (7 participants) for instance in the form of uncertainty of possible values of attributes representing physical quantities, and deviation of the expected behaviour (four participants) for instance when the effects of an event are unsure. Other definitions of uncertainty (not shown in Table 3) relate to lack of observability, internal system failures, noise, measurement uncertainty, belief uncertainty (when different stakeholders interpret the system, its elements, and its outputs, in different ways), and design uncertainty (when one is not sure about the system behavior, and hence not sure how to specify, model, and represent it).

Table 3. Definitions of uncertainty in SASs (with at least four occurrences).

| Uncertainty definitions | # |
|---|---|
| Lack of knowledge | 25 |
| Unpredictable situations | 11 |
| Property of the environment | 7 |
| Deviation from expected behavior | 4 |

> **Key findings from SQ1.2:** The results show that lack of knowledge and dealing with unpredictable situations are the most common ways to define uncertainty in self-adaptive systems. Still, several other definitions are given, showing that the opinions differ on what is uncertainty. Further, uncertainty can apply to both design time aspects or runtime aspects, and it can be internal to the system or external.

*SQ1.3: Tendency of Systems to Face Certain Sources More Frequently.*

More than half of the participants (60.3%) disagreed with the statement that self-adaptive systems tend to face certain sources of uncertainty more frequently than other. The remaining participants (39.7%) agree with the statement. Participants who agreed with the statement, listed a variety of sources of uncertainty in their explanations. The most frequently mentioned source of uncertainty in self-adaptive systems is the environment (12 times mentioned) together with internal operations and errors (8 times). The environment as a source of uncertainty mainly refers to changes in the operational context of the self-adaptive system. Other uncertainty sources mentioned more than once are monitoring and sensing (4 times) and interaction of humans with the system (2 times).

> **Key findings from SQ1.3:** The results show that there is no agreement in the community about whether self-adaptive systems face certain sources of uncertainty more frequently as others. For those that believe that self-adaptive systems tend to encounter certain sources more often, uncertainty in the environment and internal operations and errors are mentioned as the most common sources.

*4.1.2 Mitigating Uncertainty.* We formulated three survey questions to collect data for answering the second research question that focuses on how uncertainty is mitigated and how specific sources of uncertainty are handled in concrete self-adaptive systems, see Table 4.

*SQ2.1: Tackling Uncertainty at Design Time vs. Runtime.*

Almost all participants (i.e., 52) agreed that uncertainty in self-adaptive systems should be addressed both at design and runtime. Figure 5 summarises the motivations of these participants (51, i.e.,

Table 4. Survey questions for RQ2.

| ID | SQs for RQ2 focuses on mitigating uncertainty |
| --- | --- |
| SQ2.1 | Do you think uncertainty should be tamed at design-time or run-time? |
| SQ2.2 | Do you apply particular approaches to mitigate uncertainty? If yes, please explain what are the sources of uncertainty and the mitigation approach(es) you apply. |
| SQ2.3 | Continuing on your answer to the previous question, are these sources of uncertainty addressed in a systematic way (i.e., using a structured well-defined approach) or ad-hoc (i.e., on a case by case basis)? If your answer is 'systematic', give one or two examples. If your answer is 'ad-hoc' please explain whether and how this needs to be improved? |
| SQ2.4 | Consider a scenario in which a SAS needs to deal with several SoU. When dealing with such multiple SoU would you prioritise them? If yes, please select an option or explain your answer. |

96%), while Table 5 summarises a selection of codes we extracted from the answers of participants to SQ2.1, illustrated with sample quotes. Note that due to lack of a clear extended response, one answer was excluded from these statistics. Thirty-three participants (64%) motivated their response by stating that at design time strategies need to be developed that partially address uncertainty, while at runtime the system should use these strategies to take action and address the remaining uncertainties during operation. The next common justification for handling uncertainty both at design and runtime used by 5 participants (10%) is that uncertainty is continuously present over the entire life cycle of the system and therefore requires attention in both phases. Four participants (8%) stated that although uncertainty should be tamed in both phases, the strategy to be used depends on the type of uncertainty. Three respondents (6%) argued uncertainty cannot be fully tackled at design time, yet, postponing everything to run time becomes time consuming and costly. The remaining 6 participants provided more specific arguments why uncertainty needs to be tackled across the life cycle. In the discussion section we further explore the implications of the findings of SQ2.1 by deriving an initial process for uncertainty handling that spans design time and runtime.

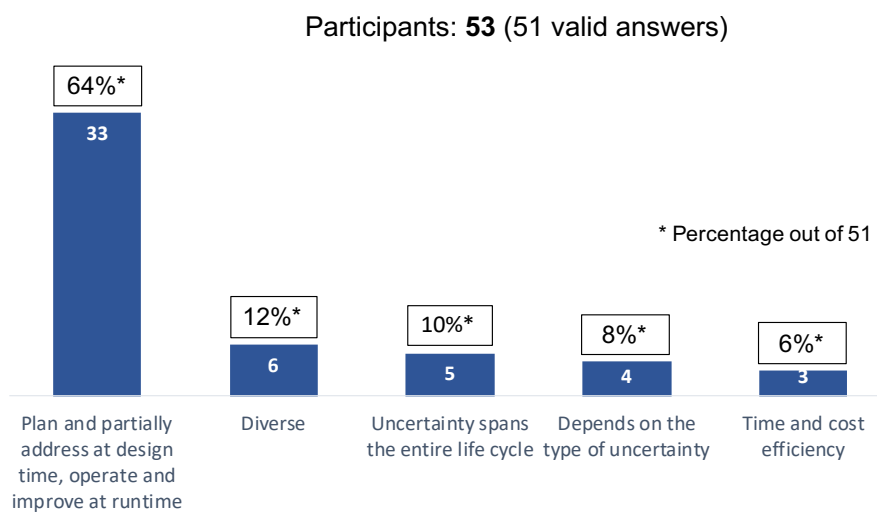

Fig. 5. Motivations for why uncertainty needs to be addressed at design and runtime.

> **Key findings from SQ2.1:** There is consensus in the community that uncertainty needs to be addressed both at design and runtime. The motivations are: a) If possible, uncertainty and/or its sources should be identified at design-time to make preparations for run-time activities, b) Strategies, recipes or templates to handle uncertainty can be created at design-time, and used at run-time, c) To decrease the costs and improve the efficiency of the system, uncertainty should be dealt with at design time as much as possible, while the rest can be postponed to runtime, d) Only when enough knowledge is not available at design time, which often is the case, uncertainty handling should be postponed to run-time.

Table 5. Qualitative analysis of SQ2.1 illustrated with sample quotes.

| Answers / Codes | Illustrative Quotes |
| --- | --- |
| Both / depends on type of uncertainty | "Some types of uncertainties can be predicted at design time, for instance, when a system is integrated in a system-of-system context and new emerging requirements are identified. However, it is possible that not all new requirements are specified at design time, so the system must be resilient enough to tackle unpredicted situations at run-time." |
| Both / time and resources demands | "Tackling uncertainty purely at runtime often takes lots of time and resources, so it should be initially tackled at design time (with all heavy calculations and modeling done online) and then the solution should be updated at runtime when the actual data from running software becomes available." |
| Both / uncertainty crosscuts lifecycle | "Since uncertainty is inherent throughout the lifecycle of a self-adaptive system, design decisions should be implemented at design-time to reduce the level of uncertainty to a reasonable level (i.e., control the controllable) while at runtime the system should be left to resolve the remaining uncertainty upon executing its task and interacting with the environment." |

### SQ2.2: Sources of Uncertainty and Associated Handling Methods.

Only seven participants (13%) provided data that actually links sources of uncertainty to uncertainty handling methods. Of these, the "environment" was the only mentioned source of uncertainty, except one that mentioned "internal operations and errors". Figure 6 shows that the most reported uncertainty handling method is the use of formal techniques (seven occurrences of a total of 20). Among these seven, four were explicitly linked the environment as source of uncertainty. The second most reported method is learning techniques (5 occurrences). None of these was explicitly linked to any type of source of uncertainty. The popularity of formal techniques is not surprisingly as these mathematical techniques can be used for a variety of tasks such as: creating/analysing models of the system and the environment, run-time analysis of alternative configurations, predicting system behaviour, and planing and decision making of adaptations. The same applies to learning methods that are used to process large volumes of data to keep runtime models updated, generate or update adaptation rules, support the analysis of models, and make predictions of the system behaviour. Sensitivity analysis with one occurrence was the only uncertainty handling method linked to internal operations and errors as source of uncertainty.

Figure 6 also highlights the distinction between specific uncertainty handling methods (shown at the top) and adaptation paradigms (shown at the bottom). Three important paradigms [34] are reported: MAPE-K based adaptation, Control Theory based, and Goal-Oriented based adaptation. Overall, the results indicate that to a certain extent, there are established practices for dealing with uncertainties, in particular those with a source in the environment. However, for other sources of uncertainty, we could not identify any established practice on how to handle them.

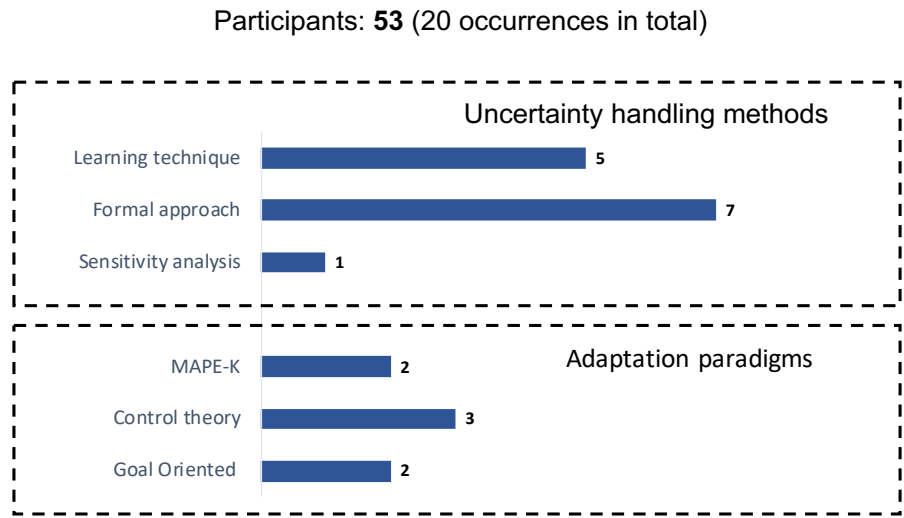

Fig. 6. Frequencies of uncertainty handling methods.

> **Key findings from SQ2.2:** Our results show that methods to handle uncertainty originating from environmental sources, which are the most common source of uncertainty, are fairly well-established. The most frequently reported methods to deal with uncertainty are formal techniques and learning, next to general paradigms to realise self-adaptation. However, these methods can be used is a broad range of tasks in self-adaptive systems, therefore their use to deal with uncertainty is not surprising.

### SQ2.3: Systematic vs. Ad-hoc Uncertainty Handling Methods.

Figure 7 shows that a majority of the respondents (29, i.e., 55%) stated that the use of systematic uncertainty handling methods depends on the types of the sources of uncertainty at hand. About one on three participants (17, i.e., 32%) stated that they use systematic uncertainty handling methods. The remaining participants (7, 13%) stated they use ad-hoc approaches and one participant did not provide an answer to this question. To get further insight in what kind of systematic uncertainty handling methods are used (by those that answered "systematic"), we asked the participants to provide examples from their practice. We identified five main classes of uncertainty handling methods, listed in Table 6.

Note that several of these "systematic" uncertainty handling approaches are in fact only partially systematic, as acknowledged by several respondents. For instance, one respondent stated "Not as systematic as I'd like, but a comprehensive use of learning and self-stabilisation seems promising. There are whole families of algorithms that fall into these categories." Note also that most systematic approaches either target design time or runtime, but do not address both phases.

The respondents that stated that the use of systematic uncertainty handling methods depends on the sources of uncertainty at hand motivate their choice with several arguments, as illustrated in Table 7. The participants that use ad-hoc uncertainty handling methods indicate that they believe that systematic methods are lacking or simply not exploited, see the examples in Table 7.

Regardless of the answer to SQ2.3, we observe that several participants acknowledge that the process of dealing with uncertainty can still be improved. We further elaborate on this insight in the discussion section, where we propose an initial uncertainty management reference process, aiming to share best practices in uncertainty management in the self-adaptive system's community.

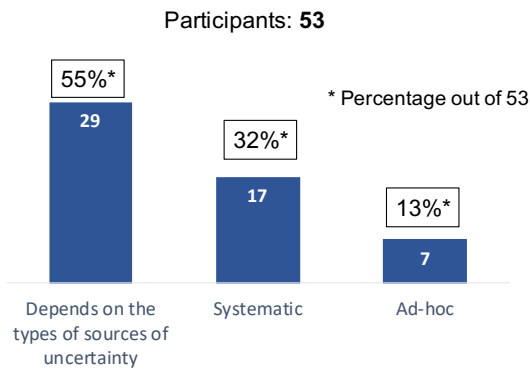

Fig. 7. Degree to which the uncertainty handling methods are structured.

Table 6. Qualitative analysis of SQ2.3 illustrated with sample quotes.

| Approaches | # | Examples |
| --- | --- | --- |
| Requirement-based | 2 | "Violations of requirements are identified and linked to reconfiguration strategies to handle uncertainty." |
| Architecture-based | 3 | "Effective software design patterns and architectural patterns are used to deal with uncertainty," "Modelling uncertainty at design-time, and systematically tracing uncertainty through the models created during development" |
| Composite-based | 2 | "Run-time adaptation within adaptation space defined at design-time, and/or run-time adaptation without knowledge of adaptation space and based on symbolic learning of suitable adaptations" |
| Monitoring-based | 4 | "Observation of self-capabilities, providing concurrent watchdogs, activating integrity fences, use of reflective architecture," "Input measurements from sensors are monitored for optimization purposes." |
| Learning-based | 2 | "Learning and self-stabilization algorithms," "Phrasing the uncertainty as a reinforcement learning problem" |

> **Key findings from SQ2.3:** The results show that more than half of the respondents stated that the use of uncertainty handling methods depends on the sources of uncertainty. Only one on three participants stated that they use systematic approaches to deal with uncertainties, with monitoring- and architecture-based methods being used most frequently. Overall, the results show that there is a need for more systematic methods to handle different types of uncertainties in self-adaptive systems.

*SQ2.4: Prioritisation of Sources of Uncertainty.*
A majority of 41 participants (i.e., 79%) stated that they use mechanisms to prioritise the order in which sources of uncertainties are handled in case of conflicts in their practice. Figure 8 shows the combinations of mechanisms that were reported at least twice (by 29 participants in total). The most frequently used mechanism (option b, 16 occurrences) gives priority to source of uncertainty with the highest impact on the overall utility of the system. The second most frequently used mechanisms give priority to source of uncertainty with the highest impact on specific non-functional requirements (option c, 4 occurrences) or they use a mechanism that takes into account also the impact on the overall utility of the system (options b and c, 4 occurrences). Twelve answers not shown in Figure 8 select different combinations of mechanisms. Notably the option that the first identified source is handled first (option a) was not selected.

Table 7. Motivations for using source-dependent uncertainty handling methods, and ad-hoc methods.

| Approaches | Example motivations |
| --- | --- |
| Depends on sources of uncertainty | "Depends on the type of the source of uncertainty but the general process is understanding the origin of the uncertainty, developing models that capture the uncertainty and it propagation within the system…," "A systematic approach would be better, but usually an ad-hoc approach is exploited," "Source of uncertainty might be the expected behaviour of a human user" |
| Ad-hoc | "It would be nice to have a checklist of things that can fail in particular type of systems (e.g. an IoT system). This would make the approach of finding what can fail more systematic." "…but common approaches could also be collected in a catalogue," "Analyzing the environmental changes are still ad-hoc and domain-specific task. However,systematic manner to analyze it is necessary," "I am not sure if there are already real standard systematic approaches for uncertainty existing," "I think systematic ways of handling different types of sources of uncertainty would be very helpful." |

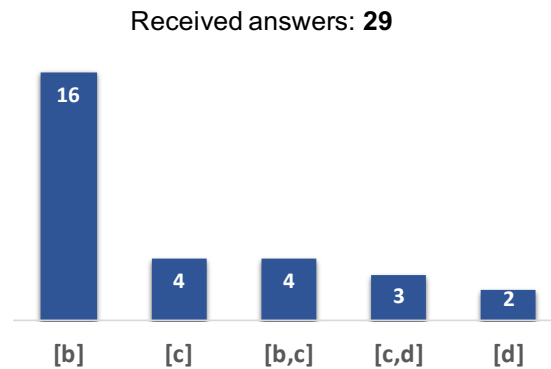

Received answers: **29**

**Options:**
a. The first identified SoU is handled first
b. SoU with a higher impact on utility gets a higher priority
c. SoU with a higher impact on specific NFRs gets a higher priority
d. SoU with a higher impact on specific system functionality gets a higher priority

Fig. 8. Prioritisation of multiple sources of uncertainty that occur concurrently.

The results show that the utility of the system is the most significant factor to decide the order in which sources of uncertainty should be addressed. In other words, the higher the impact of a source of uncertainty on the overall utility of the system, the higher its priority. The impact on non-functional requirements is also a significant factor in handling uncertainties.

> **Key findings from SQ2.4:** The results indicate that most self-adaptive systems are equipped with a prioritisation mechanism to deal with different types of uncertainties that may occur concurrently. The utility of the system is the most significant factor to decide the order in which sources of uncertainty should be addressed, next to the impact on specific non-functional requirements. Motivations for applying different prioritisation mechanisms are: a) a different operational context of the system may require a different prioritisation method, b) aspects specific to a system at hand may require specific mechanisms, c) an optimal method might in reality not be feasible and hence another method is used.

*4.1.3 Impact on Non-Functional Requirements.* We formulated two survey questions to collect data for answering the third research question that focuses on the impact of uncertainty handling methods on the non-functional requirements of self-adaptive systems, see Table 8.

Table 8. Survey Questions for RQ3.

| ID | SQs for RQ3 focusing on the effect of uncertainty on non-functional requirements |
|---|---|
| **SQ3.1** | Do you explicitly investigate how non-functional requirements of the system will be affected? If no, please motivate why not. If yes, please explain. |
| **SQ3.2** | Do you use any method to provide evidence that the system satisfies its stated functional and non-functional requirements under uncertainty? If yes, please explain. |

***SQ3.1: Impact of Uncertainty Handling Methods on Non-Functional Requirements.***
Forty-three participants (83%) stated that the systems they build explicitly take into account the effect of uncertainty on non-functional requirements. Twenty-nine among these participants provided an explanation how they achieve this, while 14 did not. We classified the techniques mentioned in the explanations in two main groups: those that consider non-functional requirements as adaptation goals (8 responses, i.e., 28%), and those that consider non-functional requirements as goals of the self-adaptive systems as a whole (21 responses, i.e., 72%), see Figure 9. Notably, most participants do not give further information about whether these techniques are applied at design time, runtime, or both (for 3 techniques we could identify whether they apply to design time, 4 apply to runtime, 6 to both, and 16 were not explained).

For the group that considers non-functional requirements as adaptation goals a variety of techniques were provided. As an example, one participant wrote: "In our case, non-functional properties (NFP)/qualities are at the core of our adaptation mechanisms, e.g., adapt the behaviour of a robot so that it achieves the best possible global balance between performance, resource consumption and safety." For the larger group that considers non-functional requirements of the self-adaptive system as a whole, we identified four main categories of techniques, as shown in Figure 9. Formal techniques and analysis techniques are the most frequently used (i.e., 21% and 10% respectively). The category "Other" (35%) refers to a variety of techniques that we could not further classify. An example is a technique that is based on correlating non-functional properties with functional properties.

The remaining 9 participants (17%) that indicated that they do not take into account the effect of uncertainty on non-functional requirements when building self-adaptive systems either stated that the topic being out of the scope in their work, or referred to obstacles they face, such as difficulty of assessing the impact on non-functional requirements. Note that one participant stated that they do not fully understand this survey question, and therefore provided no answer. This has been taken into account in the statistical analysis of SQ3.1.

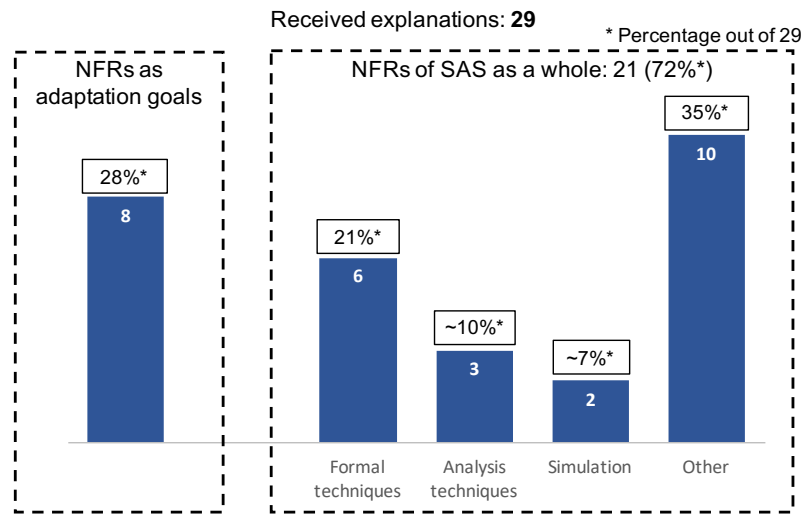

Fig. 9. Techniques for taking into account the impact of uncertainty on NFRs.

> **Key findings from SQ.3.1:** The results indicate that most existing self-adaptive systems are equipped with approaches that deal with the effects of handling uncertainty on the non-functional requirements of the system. A distinction is made between between non-functional requirements as adaptation goals and non-functional requirements of the system as a whole. For the former, no specific techniques emerged from the data. For the latter, the most popular approaches are formal techniques and analysis techniques. Difficulty of assessing the impact on non-functional requirements is a reported obstacle.

*SQ3.2 Evidence for System Compliance with Functional & Non-Functional Requirements.*

Thirty-nine participants stated that they do use some sort of approach to provide evidence that the system supports its stated functional and non-functional requirements while dealing with uncertainty. Notably most respondents did not clarify whether these methods are applied at design time, runtime or both. Thirty-two respondents provided an explanation of the method used. In Figure 10 we show that formal methods are the most commonly used approach (i.e., 37.5% of all proposed approaches) to provide evidence for supporting functional and non-functional requirements in the presence of uncertainty. The next common method is simulation (i.e., 12.5%). Run-time verification and validation, testing, and experimentation are referred to as methods used by a few of the participants (i.e., all 6.2%). The remaining methods reported by respondents (37.5%) propose other approaches to provide evidence for system compliance with requirements, such as proof of convergence, and sensitivity analysis. Two of the respondents stated that their choice of methods to provide evidence depends on the type of uncertainty they are dealing with.

> **Key findings from SQ3.2:** The results indicate that providing evidence for system compliance with non-functional requirements is common practice. The most frequently applied approach relies on formal methods. Yet a broad variety of approaches are used, ranging from simulation to testing and experimentation. While being similar with established software engineering practice, these approaches are used both at design time and runtime, but this information is not made explicit in the responses.

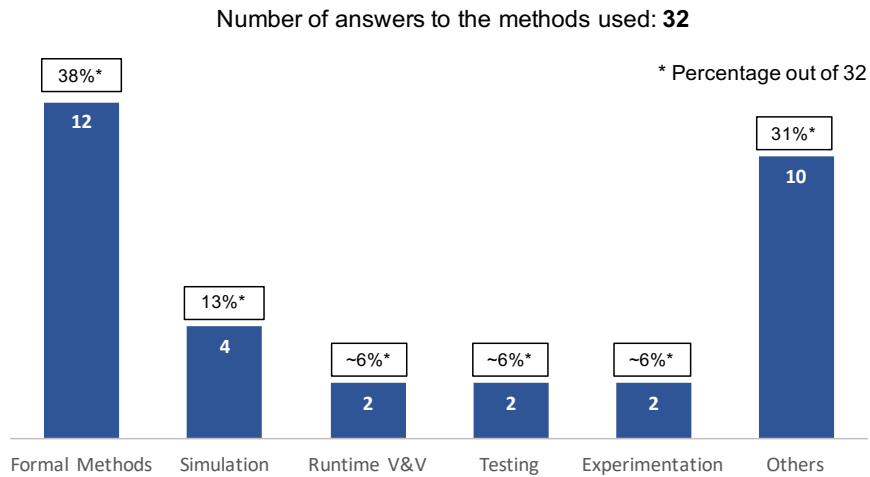

Fig. 10. Identified methods to provide evidence for system compliance with NFRs.

## 4.2 STAGE TWO - Unanticipated Change and Open Challenges of Uncertainty in SAS

For the survey of stage two, we collected data of 51 participants.[6] This data enabled us to answer the last two research questions: RQ4 looks into the ability of self-adaptive systems to deal with unanticipated changes (Subsection 4.2.1) and RQ5 looks at open challenges for research on uncertainty in self-adaptive systems with a focus on systems with strict goals (Subsection 4.2.2).

*4.2.1 Dealing with uncertainty.* We formulated two survey questions to collect data for answering the fourth research question that focuses on the ability of self-adaptive systems to deal with unanticipated changes, see Table 9.

Table 9. Survey Questions for answering RQ4.

| ID | Survey questions of RQ4 focusing on unanticipated changes |
|---|---|
| SQ4.1 | Self-adapting systems can deal only with anticipated changes. Self-adapting systems cannot deal with unanticipated changes. |
| SQ4.2 | If you selected "Disagree" as answer for Question 1, please explain how the system may be able to gain awareness of the occurrence of a change that it was not engineered to anticipate. |

### SQ4.1: Dealing with unanticipated changes.

The first question asked for the participants' view on the possibility that self-adaptive systems may only be able to deal with anticipated changes. As shown in Figure 11, only 15 of the participants (29%) held this view, with a majority of 36 participants (71%) deeming that self-adaptive systems would be able to deal with (at least some level) of unanticipated changes.

Asked to explain their position, those who considered self-adaptive systems unable to handle unanticipated changed suggested two main reasons for this: unless a change is anticipated, a system will not be able to *monitor* its occurrence, and unless a system is *built* to deal with a specific type of change from the outset, it will not be able to handle it. In contrast, the respondents who disagreed that self-adaptive systems could only deal with anticipated changes held the views that: as a matter of *principle*, self-adaptive systems ought to be able to handle unanticipated changes too; or self-adaptive systems can handle unanticipated changes *conditional* on their extent, frequency, etc. staying within certain limits; or self-adaptive systems can handle unanticipated changes as long as *learning* about them , is feasible (propositive), e.g., with human support or by employing evolutionary techniques.

---

[6]Participants were distributed as follows: 11 from SEAMS, 15 from ICAC/SASO, 11 from CASaS, and 14 extra experts.

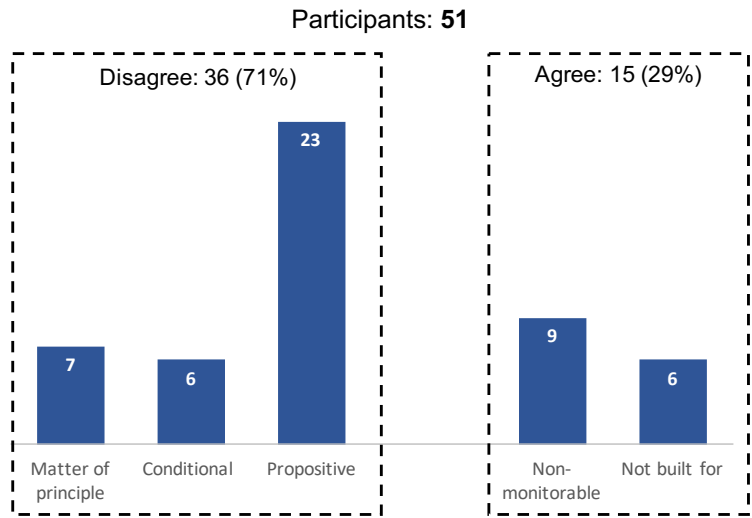

Fig. 11. Viewpoints on the ability of self-adaptive system to deal with unanticipated changes.

A broad spectrum of approaches that have the potential to allow self-adaptive systems to deal with unanticipated changes have been proposed by respondents that have the viewpoint that self-adaptive systems are able to handle unanticipated changed. These approaches ranged from software "evolution", machine learning and genetic techniques to the online synthesis (of "coping" strategies), runtime modelling, generalisation, and optimization-driven decision making. Table 10 gives a detailed overview of the qualitative analysis of SQ4.1 illustrated with example quotes.

> **Key findings from SQ4.1:** Over two thirds of the survey participants hold the view that self-adaptive systems can be engineered to cope with some level of unanticipated changes, where learning and genetic techniques have a central role. Arguments against this view are based on the inability of a self-adaptive system to monitor the occurrence of changes it was not built for or deal with such changes.

### SQ4.2: Gaining awareness of unanticipated changes.

Asking respondents that indicated that self-adaptive systems should be able to deal with unanticipated changes to suggest methods how such systems could *gain awareness* of such changes, we identified three important categories of concerns associated with handling unanticipated changes:

- Awareness – different criteria can be used to decide whether a self-adaptive system is aware of an unexpected change: simply *observing* the presence of unexpected symptoms should be regarded as awareness that an unexpected change has taken place; awareness could only be claimed when the *cause* of the unexpected change was identified by the self-adaptive system.

Table 10. Qualitative analysis of explanations from SQ4.1 with sample quotes.

| Categories & Codes | # | Example quote(s) |
|---|---|---|
| **Reasons to agree** | | |
| Not built for | 6 | "System (instances) cannot adapt to changes for which they have not been built(designed, prepared for)," "any capability to adapt to new situations where no explicit action is provided must have been built into the system" |
| Non-monitorable | 9 | "If it is not anticipated, the system cannot monitor the issue", "The possibility of change should be in some way already present in the system." |
| **Reasons to disagree** | | |
| Matter of principle | 7 | "Self-adapting system SHOULD adapt to unanticipated changes in some manner," "I disagree in principle, but I don't think we have yet reached this goal as fully as possible." |
| Conditional | 6 | "it will depend on the kind of unanticipated changes, their extent, their frequency, etc. No system will adapt to anything anytime," "It depends on what changes and reactions one wants to consider. If the reaction is always the same, then any change can be considered" |
| Propositive | 23 | "Learning approaches could allow the systems to learn new information about unanticipated changes especially if this happens with a 'human in the loop' approach", "If you can detect the consequences of the changes, you might be able to cope with that using genetic techniques" |
| **Unanticipated change support** | | |
| Evolution | 4 | "System (instances) cannot adapt to changes for which they have not been built (designed, prepared for). They require software evolution." |
| Learning | 9 | "a SAS should learn during runtime, so it should be able to deal with unexpected changes (to some extent-it depends on the knowledge base)" |
| Genetic techniques | 4 | "If you can detect the consequences of the changes, you might be able to cope with that using genetic techniques." |
| Online synthesis | 2 | "It depends on the capabilities of the system. The system may be able to recognize an unknown situation and synthetise a way to cope with it" |
| Modeling | 5 | "if unanticipated changes are reflected to model, SAS can deal with them" |
| Generalization | 2 | "Generalization capabilities of used algorithms, for example." |
| Decision Making | 2 | "the adaptation should be seen as an optimization problem and not a selection between predefined plans, No rules - mathematical optimization" |
| Others | 5 | "I imagine a system that ... discovers a new sensor and uses the input to react to changes of the environment which it was not able to detect before." |

- Identification – a range of methods for identifying unexpected changes were suggested, including: monitoring *parameter deviations* in the system parameters; observing the *consequences of changes*; noticing *mismatches* between the runtime models and runtime observations; and analyzing *historical data* collected through monitoring.
- Handling – three main classes of methods for dealing with unanticipated change were suggested. First, existing actions for dealing with expected changes can be *adapted*, e.g., by applying evolutionary approaches to the set of actions. Second, *synthesis* of new such actions, although no clear approach for this was suggested. Finally, a system can use a *default*, fail-safe action to deal with unanticipated changes, albeit in an over-conservative way.

Table 11 provides the details of the qualitative analysis of SQ4.2, illustrated with sample quotes.

Table 11. Qualitative analysis of explanations from SQ4.2 with sample quotes.

| Categories & Codes | # | Example quote(s) |
| --- | --- | --- |
| **Defining awareness** | | |
| Symptoms observed | 5 | "the change can sometimes be anticipated indirectly by affecting on other features/behaviors, i.e. [...] its partial consequence can be anticipated". |
| Cause identified | 6 | "a learning module could discover the correlation between a certain change in the environment and some bad behavior of the system and learn from this". |
| Unspecified | 6 | – |
| **Unanticipated change identification** | | |
| Monitor parameter deviations | 5 | ""drop in the system utility", "sensors are not necessary limited to detect the consequences of 'anticipated' changes" |
| Observe consequences | 5 | "multiple factors can [make] a robotic system lose its ability to make a right turn[and] it may be enough to understand the change rather than its root cause", "measuring the effect of an uncertain variable without measuring the variable directly" |
| Internal model mismatch | 5 | "having a model [...] and checking it; mismatch can indicate [unanticipated]change", "initial model can be partially wrong/incomplete" |
| Analyze history | 5 | "examining historical patterns among data/behaviors" |
| Unknown current status | 4 | "no matching rule in the knowledge base" |
| Identity change class | 2 | "predict 'classes' of likely changes carrying common characteristics and requirements for adaptation" |
| Human support | 1 | "a human in the loop could give the system awareness of the change" |
| **Unanticipated change reaction** | | |
| Adapt existing actions | 9 | "genetic algorithms could search plans similar to what the SAS knows, [and] apply [them] to new circumstances", "[use]cross learning [...] i.e. learn from similar systems to improve handling [of] changes" |
| Synthesize new actions | 4 | "[in] a situation for which [the SAS] has no solution, it reaches an exception state and [...] synthesizes a completely new adaptation", "engineering of systems at design time that will have the ability to autonomously and independently modify themselves [...] to successfully cope with the [unanticipated] changes at runtime" |
| Use default action | 3 | "If you can detect the consequences of the changes, you might be able to cope with that using genetic techniques." |

> **Key findings from SQ4.2:** The research community has mixed views on whether a self-adaptive system can be deemed aware of unanticipated changes when their symptoms are observed or only when the cause for these symptoms is identified. Three types of reactions to unanticipated changes were proposed: adapt existing actions, synthesise new ones, or just use a default fail-safe action.

*4.2.2 Challenges of uncertainty.* We formulated a final survey question to answer the fifth research question that focuses on the challenges of uncertainty in self-adaptive systems, see Table 12.

Table 12. Survey questions analysed to answer RQ5.

| ID | Survey question of RQ5 focusing on open challenges in uncertainty |
| --- | --- |
| **SQ.5** | Handling uncertainty in safety-critical self-adaptive systems (e.g., self-driving cars) is difficult because of remaining open challenges associated with (select all that apply): (a) Ensuring the scalability of self-adaptation, (b) Integrating machine learning (ML) into the self-adaptation, (c) Making the self-adaptation proactive, (d) Providing assurances. (e) Others (please specify). |

### SQ5: Challenges in Handling Uncertainties of Safety-critical Self-adaptive Systems.

Figure 12 shows that a majority of 44 participants (86%) selected the option "providing assurances that self-adaptive systems comply with their goals" as the top challenge in handling uncertainty in self-adaptive systems with critical goals. Yet, all proposed challenges listed in the questionnaire were deemed relevant by the participants as they were selected by more than half of them.

Twenty-five participants (49%) provided additional input to the free text box. We organised these comments into two categories: those that refined or provided additional information to one of the challenges listed among the options in the questionnaire, and comments that pointed out other distinct challenges. Figure 13 summarises the results.

The comments that refined the suggested challenges in the questionnaire relate to system *properties* that self-adaptation should guarantee, and aspects that should be included in the research of systems with assurances and using *machine learning*. The main properties mentioned in comments on system assurances were the safety, timeliness, reliability, and trust. Across the answers, seven respondents urged for caution and potential *risk* when applying self-adaptation to safety-critical systems. For instance, respondents noted that deciding actions in novel contexts is risky for safety-critical systems both for reactive and proactive decisions, that the use of machine learning hinders the correct behaviour of the system in first phases of its execution, and that machine learning complicates the computation of formal guarantees about the system behaviour.

The comments that pointed out other challenges for handling uncertainty in safety-critical self-adaptive systems are organised in three groups:
- Incomplete knowledge or lack of understanding of the environment – respondents noted that knowledge is limited and that models are incomplete representations.
- Human in the loop – deciding the most appropriate granularity of control operations for human operators is a challenge. Further, respondents emphasise the need for self-explainable systems, for instance for self-adaptive systems that work in cooperation with humans, and as part of the provision of assurances generated by the system.
- Ethical and moral aspects – some decisions raise moral and ethical questions, which make the correct outcome undefined or, at least, not univocal. Therefore, these type of decisions must be taken by humans.

Table 13 provides the details of the qualitative analysis of SQ5, illustrated with sample quotes.

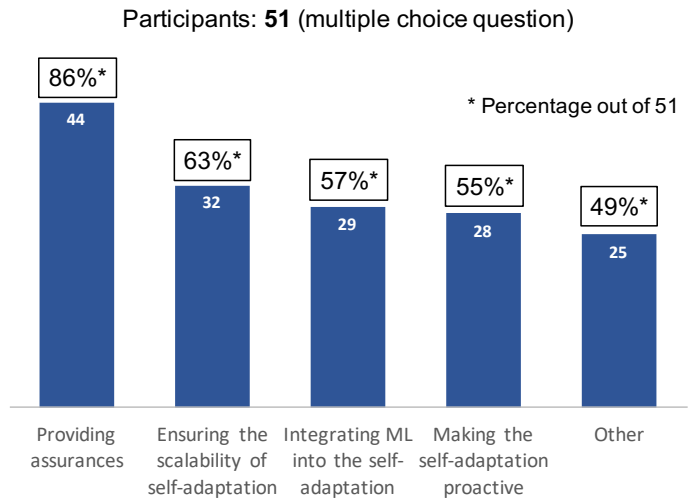

Fig. 12. Frequencies of selected challenges for handling uncertainty in safety-critical self-adaptive systems.

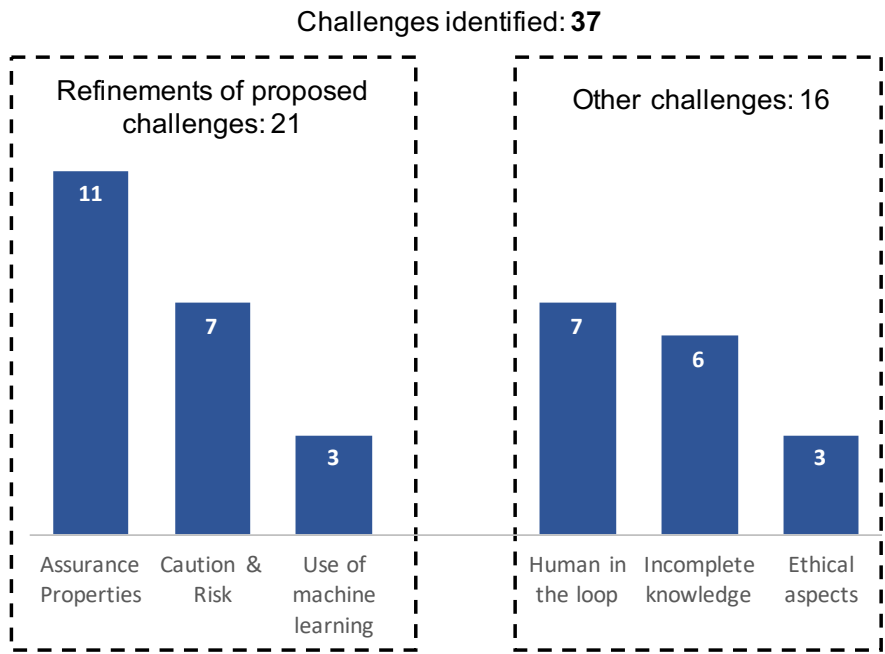

Fig. 13. Refinements of proposed challenges and other identified challenges.

> **Key findings from SQ5:** A large part of the community deems the assurances guarantee (e.g., for safety, timeliness, reliability) as a key challenge for safety-critical self-adapting systems. Caution should be taken when applying self-adaptation in safety-critical systems, in particular regarding novel situations and when machine learning is used. Dealing with lack of knowledge, supporting humans in the loop, and dealing with ethical aspects are key challenges of safety-critical self-adapting systems.

## 5 DISCUSSION

In this section we first reflect on the outcome of the analysis from Section 4, and based on that we discuss opportunities for further study. Then, we consolidate the results of the survey by proposing a reference process for uncertainty management. This process can be applied to self-adaptive systems facing similar sources of uncertainty in different application domains.

Table 13. Qualitative analysis of SQ5 with sample quotes.

| Categories & Codes | # | Example quote(s) |
|---|---|---|
| **Refinements of proposed challenges** | | |
| Assurance properties | 11 | "Ensuring the efficiency and reliability of the self-adaptation", "Providing assurances includes 3 aspects: timely assurances (in time for action), reachability/selection of solution in time (guarantee that a safety critical adaptation will converge), explanation of adaptation sufficient for the understanding+trust(collaborating agents, humans or machines)". |
| Use of machine learning | 3 | "The real challenge associated with ML here is to provide formal guarantees". |
| Caution and risk | 7 | "Both reactive and proactive decisions can be risky in novel contexts", "[with ML] the system will not behave correctly in these initial phases". |
| **Other challenges** | | |
| Incomplete knowledge | 6 | "Lack of knowledge on how to handle it, i.e., incomplete model", "Lack of knowledge about environment", "Incompleteness of the knowledge and consequently of the models representing the knowledge". |
| Human in the loop | 7 | "Providing the right granularity, API, etc. for the human control is hard", "[...] self-explainablity of the system and the role of humans –that is even if the system does something super smart, it works in cooperation with humans which have no clue what the system does and why. |
| Ethical aspects | 3 | "Moral and ethical questions which are only to be answered by humans (e.g., the well-known dilemma about risking either the life of the driver of those of passer-by in a self-driving car", "Ethical reasons: what is the ground truth (correct outcome)?" |

## 5.1 Reflections and Opportunities for Further Study

***The scope of non-functional requirements.*** One interesting aspect of the methods for handling uncertainty in self-adaptive systems is their effect on non-functional requirements. Two viewpoints can be distinguished: the effects on non-functional requirements that are the goals of adaptation, and side effects on non-functional requirements that are not goals of adaptation. Our study indicates that participants often do not distinguish between these two important viewpoints. This aspect is worth further investigation, as both viewpoints should be taken into account in the different stages of self-adaptation, from monitoring to decision-making and adapting the managed system.

***Dealing with concurrent sources of uncertainty.*** Self-adaptive systems may face several sources of uncertainty concurrently. Different types of uncertainty may have distinctive effects on different non-functional requirements. The current practice either focuses on prioritising one particular non-functional requirement, or applies best effort based on an overall utility measure for the system. For systems with critical goals, it is worth examining whether learning methods could be used to train systems facing multiple sources of uncertainty how to switch dynamically between pursuing high priority requirements and overall system utility, depending on the changing operating conditions.

***Consolidating knowledge on the concept of uncertainty.*** While this study indicates that there is growing consensus on the importance of uncertainty in relation to self-adaptation, there is still a gap in the understanding of what constitutes uncertainty and the sources of uncertainty. We argue that, even though there is a level of agreement on the foundations of the notion of uncertainty (i.e., lack of knowledge and unanticipated changes), it is essential to consolidate the knowledge on uncertainty and its sources in a standardised format; this can facilitate efficient uncertainty management and support reuse of best practices among researchers and practitioners.

***Guidelines for uncertainty handling methods.*** This study shows that a wide variety of methods are used to tame uncertainty. Yet, a coherent overview of the existing uncertainty handling methods, including application guidelines for practitioners, is currently missing. To address this need, we propose the creation of a handbook of existing uncertainty handling methods dealing with different sources of uncertainty, accompanied by guidelines on how to use them in practice. Such a handbook could categorise the available methods according to how they handle different uncertainty sources, or based on their domain of application. Application scenarios for each method may be elaborated to further help researchers and practitioners in choosing suitable methods for a problem at hand.

***Perpetual uncertainty management throughout the system lifetime.*** Sources of uncertainty are typically identified and partially handled at design-time, and further resolved at run-time. However, sources of uncertainty evolve during the lifetime of a system, and therefore identification and monitoring of sources of uncertainty should not be limited to the design-time phase. Sources of uncertainty may disappear and new ones may appear due to changes, and this may affect system functions in unpredictable ways. Hence, it is important to continuously monitor sources of uncertainty, analyse them, predict how they may change over time, and how they may affect the system behaviour. Systematic methods for managing sources of uncertainty would facilitate the modification of the uncertainty management mechanism throughout the lifetime of the system.

***Variability in software systems versus variability as a source of uncertainty.*** By definition, variability refers to anticipated changes in a software system or its environment; therefore it is predictable, and the system does not necessarily require self-adaptive capabilities to deal with variability [10]. Uncertainty in self-adaptive systems refers to deviations of deterministic knowledge that may reduce the confidence in the runtime adaptation decisions made based on that knowledge [34]. Self-adaptation exploits variability to resolve uncertainty. Our study has revealed a lack of consensus on the difference and relationship between the concepts of variability and uncertainty. This disparity is worth further investigation to define coherent terminology and to enable the development of innovative approaches for mitigating uncertainty by exploiting the broad existing knowledge on dealing with variability.

***Principled discussion on unanticipated changes.*** While our study indicates that a majority of the community agrees that self-adaptive systems can deal with unanticipated change, the ability or inability of self-adaptive systems to handle unanticipated change is subject of debate. The community would benefit from a principled discussion on this topic. This would improve our understanding of uncertainty, set the right expectations for what self-adaptive systems can handle and what is beyond their capabilities, and provide a basis for future research into the challenging problems surrounding uncertainty management in self-adaptive systems.

***Dealing with unanticipated change.*** Our study has identified a rich palette of interesting approaches for equipping self-adaptive systems with support for handling unanticipated change. The approaches put forward include integrating adaptation with evolution, and exploiting online synthesis. Nevertheless, turning these approaches into practice poses major challenges. The most promising approach may lay in exploiting machine learning and evolutionary computation techniques such as genetic algorithms. There is a strong belief that these approaches will push the abilities of self-adaptation beyond what we have been able to achieve so far. Yet, here too, there is no free lunch, as exemplified by the challenges of providing formal guarantees of correctness, timeliness, safety, etc. associated with learning techniques. Addressing the different challenges associated with these groups of methods will require substantial research effort.

## 5.2 Towards a Reference Process for Uncertainty Management in SAS

The results of this study, and particularly the comments provided by experts, make clear that there is a lack of clear guidelines and reusable artifacts from the current best practices in managing uncertainty in self-adaptive systems. The participants in our study suggested that such guidelines and knowledge would enhance the effectiveness and reliability of uncertainty management in self-adaptive systems. To address this need identified by the study, we propose a reference process for uncertainty management in self-adaptive systems. Figure 14 summarises the three-step approach that we followed to create this reference process.

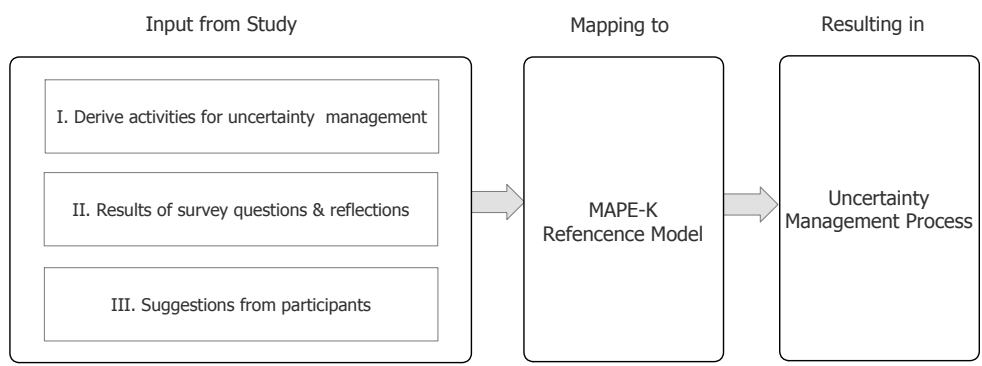

Fig. 14. Three-step approach used to derive the reference process for uncertainty management.

In the first step, we started by extracting common design and runtime activities performed to handle uncertainty in self-adaptive systems. To that end, we analysed and derived such activities from the examples given by the study participants (Input from Study I, Figure 14). Next, we used the results of the analysis of survey questions and our reflections from Section 5.1 to better understand common uncertainty handling practices, their specifications, strengths, shortcomings and areas for improvement (Input from Study II). Finally, we used the suggestions that the participants provided to the open questions, in particular their suggestions on how the current practice of handling uncertainty can be improved (Input from Study III). The result of the first stage is a set of design-time and runtime activities required for uncertainty management. We summarise this set of activities in Table 14.

In the second step of our approach, we mapped the uncertainty management activities from Table 14 to the different elements of the MAPE-K reference model. Table 15 presents this mapping, which shows clearly that handling uncertainty in self-adaptive systems involves design-time and runtime activities that affect every single element of the MAPE-K loop.

Finally, in the third step, we used our mapping of activities to MAPE-K elements to define the reference process for uncertainty management. As shown in Figure 15, this reference process groups the design-time activities into four groups:

Table 14. Synthesis of uncertainty management activities derived from participants' answers.

| ID | Activity | Description |
| --- | --- | --- |
| **Design-time** | | |
| DTA1 | Identifying uncertainty | Identifying different uncertainties the system is exposed in terms of nature, location, level, emerging time, and sources. |
| DTA2 | Identify changes of uncertainty | Pinpoint the sources of uncertainty that are prone to change over time and need to be monitored throughout the system lifetime. |
| DTA3 | Identify impact of uncertainty on goals | Determine the impact of uncertainty on non-functional requirements (both targeted by adaptation and affected by it); prioritise sources of uncertainty accordingly. |
| DTA4 | Model uncertainty | Apply modelling techniques to represent uncertainty in models of the system and its environment that will be employed at runtime. |
| DTA5 | Select uncertainty resolution techniques | Select or devise proper techniques to resolve uncertainties and deal with trade-offs to address them. |
| DTA6 | Analyse the impact of uncertainty | Apply analysis techniques to evaluate the impact of uncertainty on system behaviour and adaptation goals (e.g., cost, benefit, quality assurances, etc.); revise resolution techniques accordingly. |
| DTA7 | Devise uncertainty monitors | Select or devise monitors to track uncertainty. |
| DTA8 | Implement solutions | Realise the components for uncertainty management and deploy them. |
| **Run-time** | | |
| RTA1 | Track uncertainty propagation | Monitor the sources of uncertainty; process the data (possibly using learning or search-based techniques) and update runtime models accordingly. |
| RTA2 | Analyse the impact of uncertainty | Perform analysis to determine the impact of uncertainty on the realisation of adaptation goals (possibly using learning or search based techniques). |
| RTA3 | Assess impact of uncertainty on planning | Select alternative configurations if required and plan for adapting the managed system, assess the impact of uncertainty; take action if required (e.g., re-plan if sudden change occurs). |
| RTA4 | Assess impact of uncertainty on execution | Assess the impact of relevant sources of uncertainty on the effects of adaptation actions; take action if required (e.g., adjust the workflow of adaptation actions). |

(1) The *Identify uncertainty* activity group includes the activities required to identify the uncertainties that the system is exposed to (activity DTA1 from Table 14), the uncertainty sources prone to change over time (DTA2), and the impact of uncertainty on non-functional requirements (DTA3).
(2) The *Model uncertainty* activity group comprises the design-time modelling of uncertainty (DTA4), and the selection/devising of uncertainty resolution techniques (DTA5).

Table 15. Mapping of uncertainty management activities to MAPE-K elements.

| Activity | Monitor | Analyze | Plan | Execute | Knowledge | MAPE-K |
|----------|---------|---------|------|---------|-----------|--------|
| **DTA1** |         |         |      |         |           | X      |
| **DTA2** | X       |         |      |         |           |        |
| **DTA3** | X       |         |      |         |           |        |
| **DTA4** |         |         |      |         | X         |        |
| **DTA5** |         | X       | X    |         |           |        |
| **DTA6** |         |         |      |         |           | X      |
| **DTA7** | X       |         |      |         |           |        |
| **DTA8** |         |         |      |         |           | X      |
| **RTA1** | X       |         |      |         | X         |        |
| **RTA2** |         | X       |      |         | X         |        |
| **RTA3** |         |         | X    |         | X         |        |
| **RTA4** |         |         |      | X       | X         |        |

(3) The *Analyse impact of uncertainty* activity group involves the application of techniques for evaluating the impact that uncertainty may have on the self-adaptive system and its adaptation goals (DTA6).
(4) The *Implement uncertainty handling mechanisms* activity group consists of activities for selecting or devising uncertainty-tracking monitors (DTA7) and for implementing the MAPE-K components required for uncertainty management (DTA8).

Performing these design-time activities produces an *Uncertainty Blueprint* that comprises five core components:

(1) *Uncertainty sources* are measurable properties of the system, the environment, or goals that may affect the behaviour of the self-adaptive system. An example of an uncertainty source in the environment is interference along the links of a wireless network of an IoT system.
(2) *Uncertainty sensors* are means to measure the sources of uncertainty, either in the environment, the system or the goals of the system. An example is a sensor that tracks the actual level of interference along the network links of an IoT system.
(3) *Uncertainty-aware runtime models* are runtime abstractions of the system or any aspect related to the system that represents uncertainty as first-class citizen; these models are kept updated during operation and used for the decision-making of adaptation. An example is an automata model that represents the reliability of an IoT system where links have parameters that represent probabilities of packet loss.
(4) *Uncertainty resolution techniques* are approaches that enable the analysis of alternatives configurations taken into account the historical, measured, or predicted uncertainty. An example is a runtime verification of an automata model of an IoT system to determine the expected packet loss based on up to data data of interference along links.
(5) *Uncertainty assessment techniques* are approaches that enable to assess the impact of decision-making on the planning and execution of system adaptations. An example is a a runtime component that tracks the effectiveness of updated network settings of an IoT system applied using adaptation.

These components are then used to realise the following runtime activities, which are required for mitigating uncertainty during operation:

(1) *Track uncertainty propagation* uses the knowledge about uncertainty sources and the uncertainty sensors to monitor the sources of uncertainty and to update the runtime models of the self-adaptive system in line with new information obtained about these sources.
(2) *Analyse the impact of uncertainty* uses the uncertainty-aware runtime models and the uncertainty resolution techniques to determine the impact of uncertainty on the realisation of the goals of the self-adaptive system.
(3) *Assess the impact of uncertainty on planning* uses uncertainty assessment techniques to support planning under uncertainty.
(4) *Assess the impact of uncertainty on execution* uses uncertainty assessment techniques (like the previous runtime activity), with a focus on identifying how uncertainty may impact the effect of adaptation actions, and on reacting to mitigate this impact if needed.

The reference process for uncertainty management consolidates the current common practices derived from our study into a coherent format. As such, it advances the knowledge on uncertainty mitigation in self-adaptive systems, and can support practitioners in pursuing a more systematic, as well as time- and cost-efficient approach to managing uncertainty in self-adaptive systems. We kept the reference process generic on purpose, to enable its instantiation at different levels of abstraction, such as a particular application domain, a family of systems (e.g. a software product line), or even a single system. We foresee that such instantiation will yield reusable artifacts, both at the level of the process itself, and at the level of the components of the *Uncertainty Blueprint*.

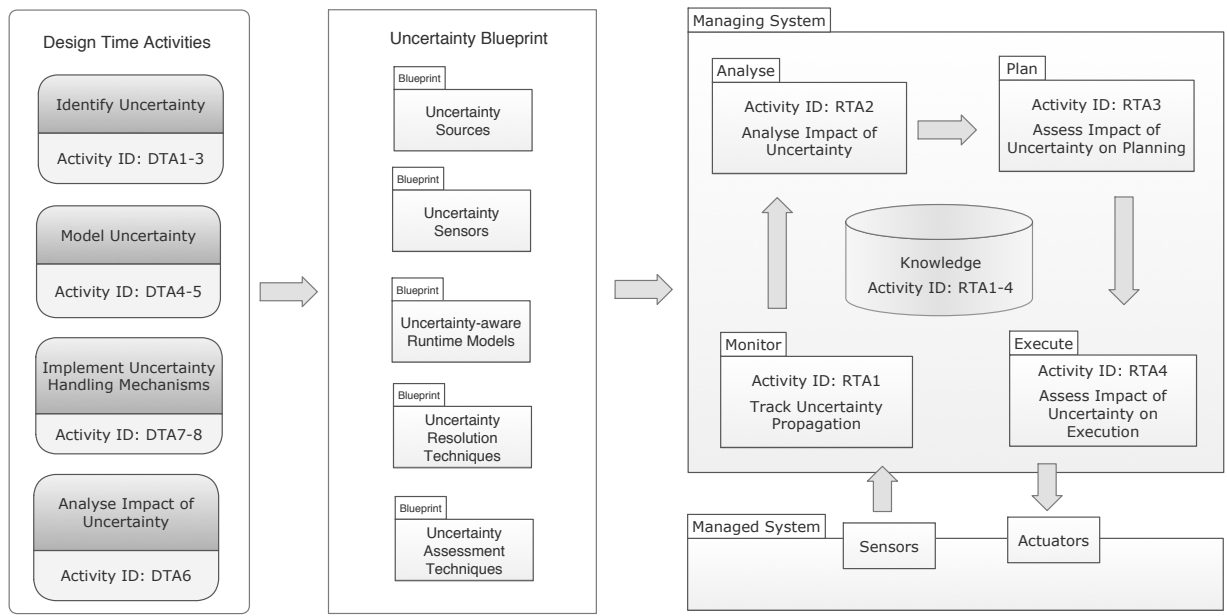

Fig. 15.  - A reference process for uncertainty management throughout the lifecycle of the system.

## 6 THREATS TO VALIDITY

We use the guidelines proposed in [36] to assess threats to the validity of this study. We focus on construct validity (extent to which we obtained the right measure and whether we defined the right scope for the study goal), external validity (extent to which the findings can be generalized), and reliability (extent to which we can ensure that our results are the same if our study is done again).

## 6.1 Construct Validity

The core construct in our surveys are knowledge of self-adaptive systems, the definition of uncertainty and its sources. Our results suggest that a common understanding of these concepts exists. To further reduce possible misinterpretations, we took two actions for stage one survey. First, the questionnaire included mainly close-ended questions (i.e., 8 questions out of 9) combined with open-ended sub-questions. Having close-ended questions reduces the probability of misinterpretations, and helps the participants to better understand the question, while the open-ended part offers them the freedom to express their opinion. Second, we conducted a pilot in which we used the feedback of several participants to enhance the comprehensible and clarity of the questions. We mitigated this threat in the second stage survey by selecting experienced participants at the main venues of the community and invited additional experts ensuring that the required basic knowledge was present. Additionally, respondents could clarify issues in the free text provided with the questions. Some questions may have been formulated such that respondents were forced to provide an answer that may not have objectively expressed their opinion. We mitigated this threat by allowing the respondents to provide comments with their answers. To mitigate bias in the formulation of the questions, we used a refinement process when defining the questions, where four researchers reviewed the questions, individually and as a team.

## 6.2 External Validity

Generalization of the study results might be a potential threat to validity. The main issue here is the selection of the sample of the population may not have been representative. This may lead to study results that may be imprecise. When using non-probabilistic sampling method, there is always a risk that the sample used to conduct the survey is biased and not representative of the target population. To mitigate the external validity threat we decided to target and reach out to experts included in the SEAMS mailing list for the first stage survey. This was our best chance for connecting with subjects within the target population. To ensure that participants have the required hands-on experience, we included several open-ended questions asking about their personal experience while dealing with uncertainty in practice, and requested for real life examples of systems they have worked on in the past. Based on their responses to open-ended questions, we were able to get an understanding of their expertise and filter out inapplicable responses. Similarly, to resolve this problem during the stage two survey, we selected participants at the main venues of the community and invited additional experts, increasing the confidence that the sample was representative.

## 6.3 Reliability

Data analysis and coding in particular are creative tasks that are to some extent subjective. To mitigate interpretation bias, we followed several strategies. In the stage one survey one author went through the data independently, and discussed with a second researchers where needed until an agreement was reached. In the second survey two researchers performed the data analysis of each question in an iterative way and then the results were cross-checked by the two other researchers, any differences where discussed until we reached consensus, In addition, were applicable, we formatted the questionnaire in a preventive way, such that it would mitigate the responses' susceptibility to multiple interpretations. To realise this, we designed most of our questions in close-ended and multiple-choice format. In case of open-ended questions, where applicable we consolidated these questions with close-ended questions (i.e., "Yes" or "No", and "Agree" or "disagree") as well, which ultimately clarified the answers and helped us to better understand the free text responses. In addition, we made all the material of the survey publicly available.

## 7 CONCLUSION

We presented an extensive study into the perception of the community on the concept of uncertainty, approaches to handle uncertainty, and open challenges in this area. To that end, we devised a two-stage research approach, each stage involving a survey with a distinct focus. The study results yielded a variety of consolidated insights, including an overview of uncertainty sources considered in self-adaptive systems, and an overview of existing uncertainty handling methods used in the development of self-adaptive systems.

The results also highlight aspects of uncertainty for which consensus exists in the community. These aspects include the community's views on what constitutes uncertainty (lack of knowledge and unpredictable situations), the fact that uncertainty needs to be addressed both at design time and runtime, the importance and common use of prioritisation mechanisms for dealing with different types of uncertainties that may occur concurrently, the importance of providing evidence for system compliance with non-functional requirements (mostly relying on formal techniques), and the importance of assurance guarantees as a key challenge for safety-critical self-adaptive systems. On the other hand, the study reveals multiple aspects for which no consensus exists. Among these are mixed opinions on whether uncertainty is the essential motivation for applying self-adaptation. Furthermore, while there is a widespread belief that self-adaptive systems can be engineered to cope with some level of unanticipated changes, there are mixed views on whether a self-adaptive system can be deemed aware of such unanticipated changes.

Finally, this study revealed the lack of systematic approaches for managing uncertainty. To address this gap, we presented an initial reference model for managing uncertainty in self-adaptive systems. This reusable process builds upon and consolidates the results of our study. We hope that researchers and engineers will find the process useful to improve the way they manage uncertainty when developing self-adaptive systems.